\documentclass[a4paper,twocolumn,english,aps,prl,superscriptaddress]{revtex4-1}
\usepackage[T1]{fontenc}
\usepackage[latin9]{inputenc}
\usepackage{babel}
\usepackage{amsmath}
\usepackage{amssymb}
\usepackage{graphicx}
\usepackage[unicode=true,pdfusetitle,
 bookmarks=true,bookmarksnumbered=false,bookmarksopen=false,
 breaklinks=false,pdfborder={0 0 1},backref=false,colorlinks=false]
 {hyperref}
\usepackage{breakurl}

\makeatletter

\usepackage{babel}

\makeatother

\begin{document}

\title{Majorana-Josephson Interferometer}

\author{Chang-An Li }

\affiliation{Department of Physics, The University of Hong Kong, Pokfulam Road,
Hong Kong, China }

\author{Jian Li }
\email{lijian@westlake.edu.cn}

\affiliation{Institute of Natural Sciences, Westlake Institute for Advanced Study,
Hangzhou, Zhejiang, China}

\affiliation{School of Science, Westlake University, Hangzhou, Zhejiang, China}

\author{Shun-Qing Shen }
\email{sshen@hku.hk}

\affiliation{Department of Physics, The University of Hong Kong, Pokfulam Road,
Hong Kong, China }

\affiliation{Kavli Institute for Theoretical Sciences, University of Chinese Academy
of Sciences, Beijing, China}

\date{\today }
\begin{abstract}
We propose an interferometer for chiral Majorana modes where the interference
effect is caused and controlled by a Josephson junction of proximity-induced
topological superconductors, hence, a Majorana-Josephson interferometer.
This interferometer is based on a two-terminal quantum anomalous Hall
bar, and as such its transport observables exhibit interference patterns
depending on both the Josephson phase and the junction length. Observing
these interference patterns will establish quantum coherent Majorana
transport and further provide a powerful characterization tool for
the relevant system.
\end{abstract}
\maketitle
\textit{Introduction.}\textemdash The emergence of Majorana fermion
modes in condensed matter systems \cite{Read00prb,Kitaev01usp,FuL08prl,Oreg10prl,Lutchyn10prl,Wilczek09np,Alicea12rpp,Beenakker13arcmp,Elliott-15rmp,Sato17rpp,SQS}
has shed light on the feasible realization of topological quantum computation
\cite{Kitaev03ap,Nayak08rmp,Alicea11np,Stern13science,DasSarma15npj,Karzig17prb,Obrien18prl,Litinski18prb}.
To this day, observations of Majorana modes have been reported in
various structures exhibiting topological superconductivity \cite{Mourik12science,Williams12prl,DengMT12nanoletter,Das12np,Nadj14science,XuJP15prl,Jeon17science,He17sci,ZhangP18science,ZhangH18nature,Bernevigbook}.
It becomes imperative to demonstrate the quantum coherent manipulation
of Majorana modes in order, for example, to showcase the much desired
non-Abelian braiding statistics \cite{Moore91npb,WenXG91prl,Ivanov01prl,Stern10nature}.
One appealing route towards this goal involves the utilization of
interferometers, which was originally proposed for the fractional
quantum Hall anyonic platform \cite{DasSarma05prl,Bonderson06prl,Feldman06prl,Bishara09prb}.
Indeed, building interferometers of chiral Majorana modes ($\chi$MMs)
\cite{LawKT09prl,FuL09prl,Akhmerov09prl,Strubi11prl,LiJ12prb} can
be particularly facilitated by hybrid structures \cite{Qi10prb,ChungSB11prb,WangJ15prb}
composed of quantum anomalous Hall insulators (QAHIs) \cite{Yu10Science,Chang13sci,LuHZ13prl}
and conventional superconductors. Such a Majorana interferometer,
in turn, can serve to pinpoint the presence of quantum coherent Majorana
transport in the device \cite{He17sci}, where inelastic scattering
may otherwise obscure the current experimental evidence \cite{Ji18prl,HuangY18prb}.

\begin{figure}
\includegraphics[width=1\linewidth]{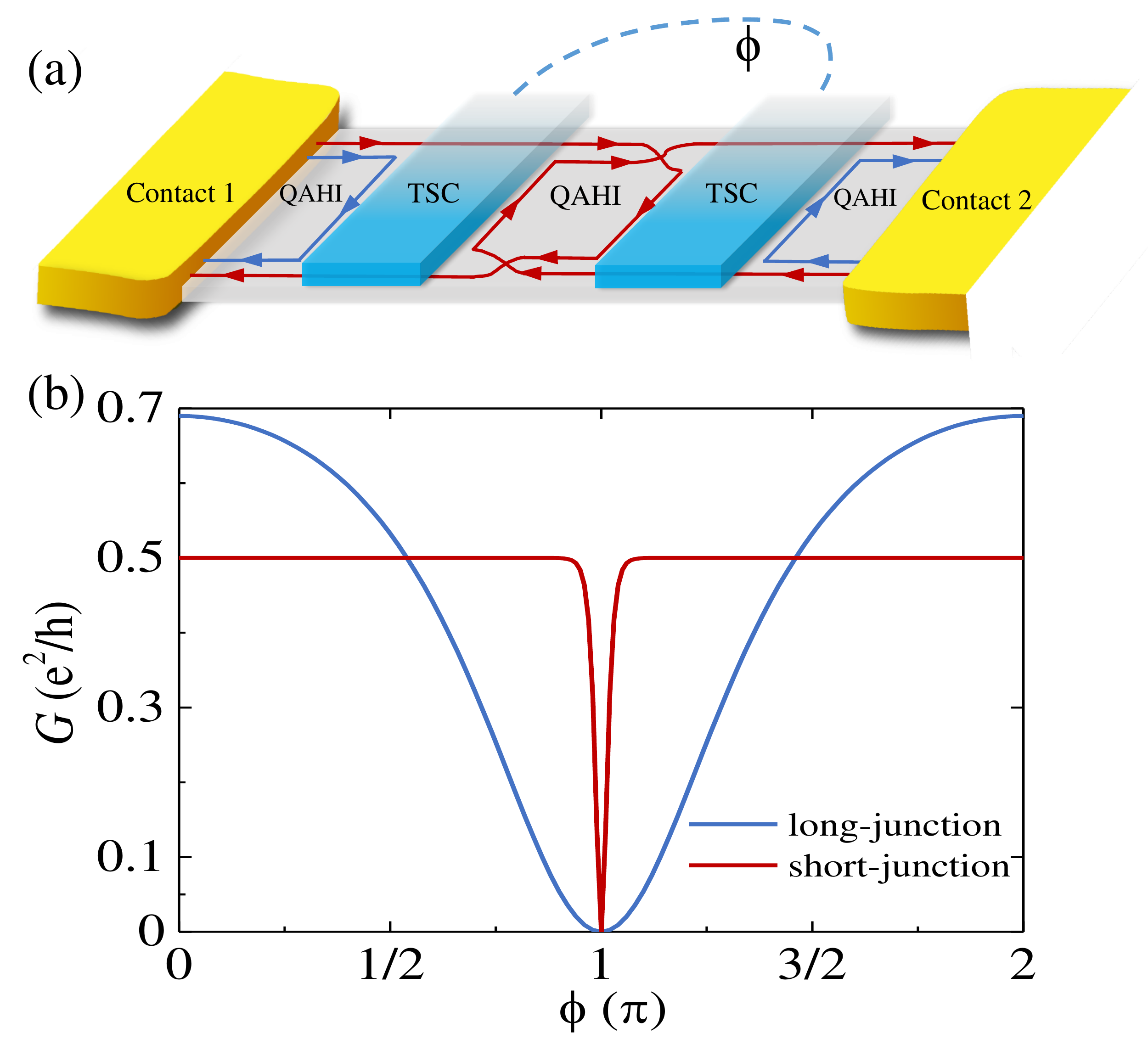}

\caption{(a) Schematic of a Majorana-Josephson interferometer. A quantum anomalous
Hall bar (gray) is in contact with two conventional superconductors
(cyan) on its top and two normal metals (yellow) at its ends. The
two superconducting contacts are grounded and maintain a tunable phase
difference $\phi$. The arrowed lines represent generic scattering
of chiral Majorana modes in the interferometer. (b) Typical two-terminal
conductance $G$ as a function $\phi$ in a Majorana-Josephson interferometer
when the superconducting junction is long (blue line) and short (red
line), respectively.\label{fig:Interfereometer}}
\end{figure}

In this paper, we propose a Majorana interferometer with its interference
loop generated and controlled by a Josephson junction, as illustrated
in Fig.~\ref{fig:Interfereometer}(a). The Josephson junction is
composed of two topological superconductors (TSCs) induced by conventional
superconductors in contact with a QAHI \cite{He17sci,Qi10prb,ChungSB11prb,WangJ15prb}. Such a Josephson junction effectively polarizes and filters a $\chi$MM in terms of its $U(1)$ degree of freedom associated with the superconducting phase. As a consequence of this novel \textit{Majorana valve effect}, quantum interference patterns in two-terminal conductance measured with the normal metallic contacts can be observed by tuning the Josephson phase $\phi$, as exemplified in Fig.~\ref{fig:Interfereometer}(b). This
Majorana-Josephson interferometer (MJI), on the one hand, extends
straightforwardly an existing experimental setup \cite{He17sci},
and hence is expected to be readily accessible. On the other hand,
its interference effect demonstrates highly nontrivial Majorana physics,
and can be used not only as a smoking-gun signature for the presence
of $\chi$MMs, but also potentially in operations of Majorana-based
topological quantum computation.

\begin{figure}
\includegraphics[width=1\linewidth]{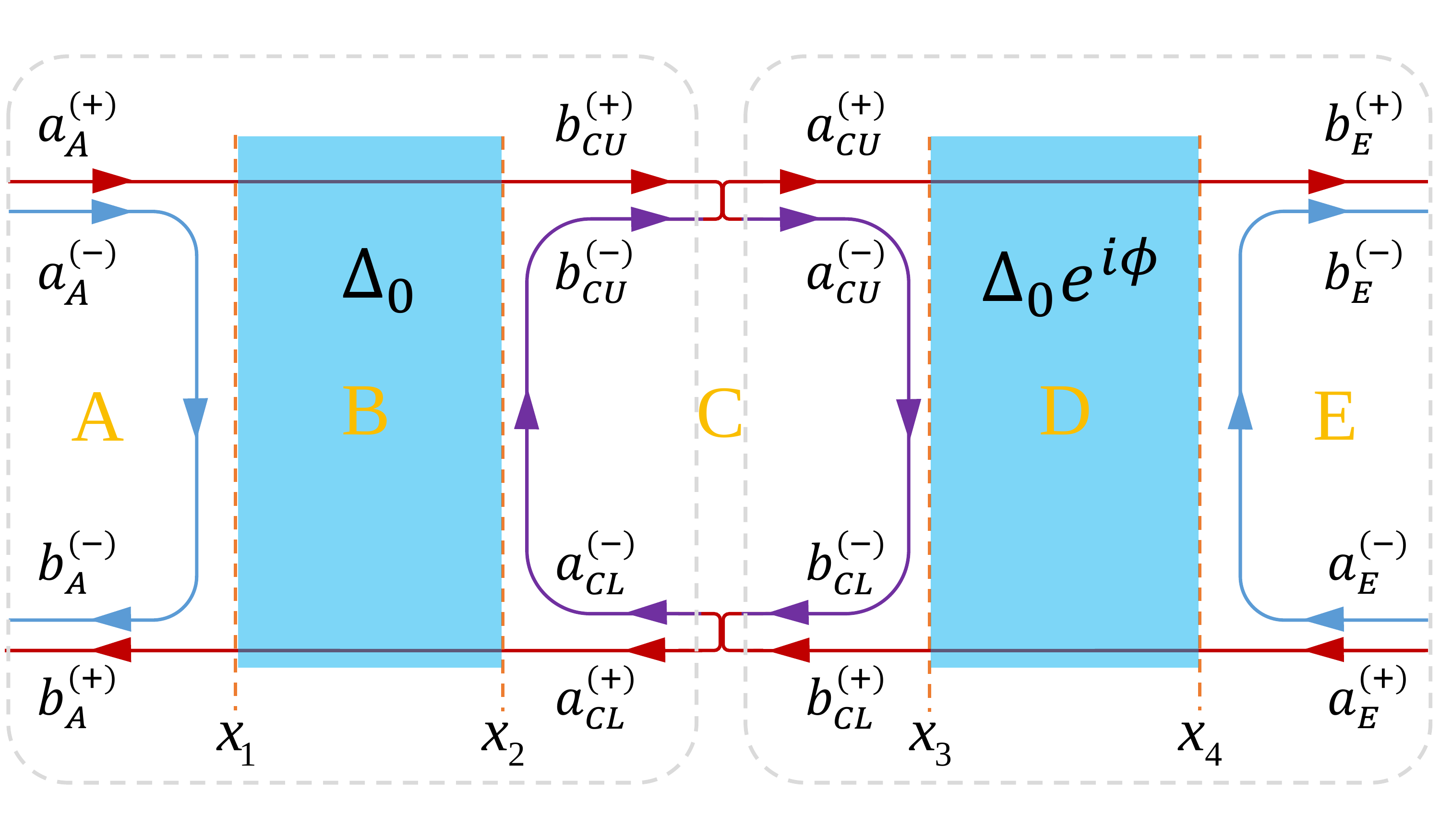}

\caption{Scattering picture of the chiral Majorana modes in the long-junction
limit of a Majorana-Josephson interferometer at $\mu=0$. Regions
with different superconducting order parameters are labeled by A through
E. The full scattering matrix can be obtained by first analyzing the
composite regions ABC and CDE (in gray dashed frames) individually,
and then connecting them by taking into account the $\phi$-dependent
local basis. Labels for the incoming ($a$'s) and the outgoing ($b$'s)
Majorana modes are indicated.\label{fig:scattering}}
\end{figure}

\textit{Model for Majorana Josephson interferometer.} \textemdash The
Bogoliubov-de Gennes Hamiltonian that describes the low-energy physics
of the MJI sample area is given by \cite{Qi10prb,ChungSB11prb}
\begin{align}
H=\begin{pmatrix}h_{0}(\hat{{\bf k}})-\mu & \Delta(x)\\
\Delta(x)^{*} & h_{0}(-\hat{{\bf k}})+\mu
\end{pmatrix},\label{eq:H_BdG}
\end{align}
where $h_{0}(\hat{{\bf k}})=(b\hat{{\bf k}}^{2}-m)\sigma_{z}+v\hat{k}_{x}\sigma_{x}+v\hat{k}_{y}\sigma_{y}$
is the effective Hamiltonian for the underlying QAHI with positive
parameters $b$, $m$ and $v$, the Pauli matrices $\sigma_{x,y,z}$
for spin, and two-dimensional wavevector operator $\hat{{\bf k}}\equiv(\hat{k}_{x},\hat{k}_{y})\equiv-i(\partial_{x},\partial_{y})$;
$\mu$ is the chemical potential. The proximity-induced pairing potential
across the sample is assumed to depend only on $x$ (see Fig.~\ref{fig:scattering}):
$\Delta(x)=\Delta_{0}$ if $x_{1}<x<x_{2}$; $\Delta_{0}e^{i\phi}$
if $x_{3}<x<x_{4}$; and $0$ otherwise, where $\Delta_{0}$ is taken
to be positive, and $\phi$ stands for the Josephson phase. In the
Hamiltonian in Eq.~\eqref{eq:H_BdG}, we have adopted the Nambu basis
which, in real space, reads $\Psi_{{\bf r}}=\left(c_{{\bf r}\uparrow},c_{{\bf r}\downarrow},c_{{\bf r}\downarrow}^{\dagger},-c_{{\bf r}\uparrow}^{\dagger}\right)^{T}$
with $c_{{\bf r}s}$ and $c_{{\bf r}s}^{\dagger}$ the annihilation
and creation operators for an electron with spin $s=\uparrow,\downarrow$
at ${\bf r}=(x,y)$, respectively. This Hamiltonian is manifestly
particle-hole symmetric: $\mathcal{P}H\mathcal{P}^{-1}=-H$ with the
particle-hole operator $\mathcal{P}=\tau_{y}\otimes\sigma_{y}\mathcal{K}$,
where $\mathcal{K}$ is the complex conjugate operator, and $\tau_{x,y,z}$
are the Pauli matrices for a Nambu spinor.

The above model defines an MJI if $\Delta_{0}^{2}>m^{2}-\mu^{2}>0$
such that, by labeling the regions with different pairing potentials
to be A to E as shown in Fig.~\ref{fig:scattering}, the topological
invariant $\mathcal{N}=2$ in the bulk of regions A, C and E, and
$\mathcal{N}=1$ in the bulk of regions B and D \cite{Qi10prb}. Throughout
this paper, we assume that the relevant energy range, determined by
temperature, bias voltages \textit{etc.}, is close enough to zero
energy, such that the scattering processes are approximately energy-independent.
We also assume that the sizes of regions A and E (B and D) are large
compared with the transverse penetration length $\xi_{\mathrm{QAH}}$
($\xi_{\mathrm{TSC}}$) of the QAHI (TSC) edge modes, such that the
scattering channels as depicted in Fig.~\ref{fig:scattering} are
well defined. We distinguish, however, two limits in terms of the
length of region C, $l_{C}\equiv x_{3}-x_{2}$, which separates the
two TSC regions: the long-junction limit with $l_{C}\gg\xi_{\mathrm{QAH}}$,
and the short-junction limit with $l_{C}\ll\xi_{\mathrm{QAH}}$.

Before we analyze the transport behaviors of the MJI, it is useful
to gain insight from the solutions of the $\chi$MMs, denoted by $\Psi_{B,D}$
in the TSC regions B and D, respectively (see Sec. I in Ref. \cite{Note-on-SM}).
At $E=0$, both solutions satisfy the Majorana condition $\mathcal{P}\Psi_{B,D}=\Psi_{B,D}$.
In addition, because the bulk Hamiltonians in regions B and D differ
only in the superconducting phase, $\Psi_{B}$ and $\Psi_{D}$ are
related by a simple transformation $\Psi_{D}=U(\phi)\Psi_{B}$ with
$U(\phi)=\exp({i\frac{\phi}{2}\tau_{z}})\otimes\sigma_{0}$. As $\mathcal{P}$
is an antiunitary operator, it follows immediately that
\begin{align}
\langle\Psi_{B}|\Psi_{D}\rangle=\langle\Psi_{D}|\Psi_{B}\rangle=\cos\frac{\phi}{2},\label{eq:mismatch}
\end{align}
which represents a mismatch between the two $\chi$MMs at finite $\phi$.
Physically, this implies an inner $U(1)$ degree of freedom associated
with the $\chi$MMs \cite{Note-on-MP}, or \textit{Majorana polarization}
as analogous to the spin polarization of spin-$\frac{1}{2}$ particles.
Thus the TSC Josephson junction effectively becomes a \textit{Majorana
valve}, similar to a spin valve \cite{Datta90apl,Dieny91prb,Zutic04rmp}
by the same analogy. This Majorana valve leads directly to an interference
loop in the MJI, as we proceed to show.

\textit{The long-junction limit.}\textemdash{} When the two TSC regions
are well separated, i.e., when $l_{C}\gg\xi_{\mathrm{QAH}}$, the
MJI can be analyzed by first considering the composite ABC region
or CDE region individually, and then treating their connection with
care. This procedure is particularly physical transparent in the $\mu=0$
case, where the partial Hamiltonian for either region ABC or CDE can
be brought to a block-diagonal form by a global unitary transformation
\cite{Qi10prb,ChungSB11prb}: $U_{p}^{\dagger}H_{p}(\mu=0)U_{p}=h_{p}^{(+)}\oplus h_{p}^{(-)}$
with $p=ABC$ or $CDE$. Here, $h_{p}^{(\pm)}=h_{0}(\hat{{\bf k}})\mp|\Delta(x)|\sigma_{z}$,
\begin{align}
 & U_{ABC}=\frac{1}{\sqrt{2}}\begin{pmatrix}\sigma_{0} & \sigma_{0}\\
-\sigma_{z} & \sigma_{z}
\end{pmatrix},\quad U_{CDE}=U(\phi)U_{ABC}.\label{eq:U_p}
\end{align}
For the sake of clarity, we assume that the partial Hamiltonian with
$p=ABC$ ($p=CDE$) is limited to the range $x<\frac{x_{2}+x_{3}}{2}$
($x>\frac{x_{2}+x_{3}}{2}$). The particle-hole operator in the transformed
basis also becomes block-diagonal and is identical for $p=ABC$ and
$p=CDE$: $\tilde{\mathcal{P}}\equiv U_{p}^{\dagger}\mathcal{P}U_{p}=-\sigma_{z}\otimes\sigma_{x}\mathcal{K}$,
which indicates that each block may allow for $\chi$MM solutions
independently. Indeed, the two subspaces corresponding to $h_{p}^{(\pm)}$
each support one $\chi$MM along the QAHI edge, but scattered differently
at the QAHI-TSC interfaces (see Fig.~\ref{fig:scattering}). This
scenario, for $p=ABC$ or $p=CDE$ individually, has been analyzed
by Chung \textit{et al.} \cite{ChungSB11prb}, and the scattering
matrix in the Majorana basis is given by
\begin{align}
\begin{pmatrix}b_{p,U}^{(+)}\\
b_{p,L}^{(+)}\\
b_{p,U}^{(-)}\\
b_{p,L}^{(-)}
\end{pmatrix}=\begin{pmatrix}1 & 0 & 0 & 0\\
0 & 1 & 0 & 0\\
0 & 0 & 0 & -1\\
0 & 0 & 1 & 0
\end{pmatrix}\begin{pmatrix}a_{p,U}^{(+)}\\
a_{p,L}^{(+)}\\
a_{p,U}^{(-)}\\
a_{p,L}^{(-)}
\end{pmatrix},\label{eq:Sp}
\end{align}
where $a_{p,U/L}^{(+/-)}$ ($b_{p,U/L}^{(+/-)}$) stands for the incoming
(outgoing) Majorana current amplitude corresponding to the $h_{p}^{(+/-)}$
block along the upper/lower edge of region $p$. Note that the $-1$
in the above scattering matrix comes from the requirement that the
determinant of the full scattering matrix is $+1$ (see Sec. II A
in Ref. \cite{Note-on-SM}). Hereafter we will abbreviate the labels
of these amplitudes according to Fig.~\ref{fig:scattering} without
ambiguity.

The key idea of the MJI proposed in this paper is that, despite the
trivial appearance of the scattering processes in either the ABC or
the CDE region individually, the connection between the two parts
is nontrivial as suggested by the Majorana polarization mismatch in
Eq.~\eqref{eq:mismatch}. The same mismatch is reflected in Eq.~\eqref{eq:U_p}
as the different basis used for $p=ABC$ and $p=CDE$ in block-diagonalizing
the partial Hamiltonians when $\phi\ne0$. It follows that the change
of basis introduces effective scattering between the $\chi$MMs as
(see Sec. II A in Ref. \cite{Note-on-SM})
\begin{align}
\begin{pmatrix}a_{C\beta}^{(+)}\\
a_{C\beta}^{(-)}
\end{pmatrix}=\begin{pmatrix}\cos\varphi_{\beta} & \sin\varphi_{\beta}\\
-\sin\varphi_{\beta} & \cos\varphi_{\beta}
\end{pmatrix}\begin{pmatrix}b_{C\beta}^{(+)}\\
b_{C\beta}^{(-)}
\end{pmatrix},\label{eq:Sc}
\end{align}
where $\beta=U,L$ and $\varphi_{U/L}=\pm\phi/2$. Combining this
equation and Eq.~\eqref{eq:Sp}, we obtain the full Majorana scattering
matrix connecting the two normal contacts to be
\begin{align}
\begin{pmatrix}b_{E}^{(+)}\\
b_{A}^{(+)}\\
b_{E}^{(-)}\\
b_{A}^{(-)}
\end{pmatrix}=\begin{pmatrix}t & -r & 0 & 0\\
r & t & 0 & 0\\
0 & 0 & 0 & -1\\
0 & 0 & 1 & 0
\end{pmatrix}\begin{pmatrix}a_{A}^{(+)}\\
a_{E}^{(+)}\\
a_{A}^{(-)}\\
a_{E}^{(-)}
\end{pmatrix},\label{eq:SM}
\end{align}
where $t=\frac{2\cos\frac{\phi}{2}}{1+\cos^{2}\frac{\phi}{2}}$ and
$r=\frac{1-\cos^{2}\frac{\phi}{2}}{1+\cos^{2}\frac{\phi}{2}}$. Thus
the MJI functions effectively as a Fabry-Perot interferometer for
$\chi$MMs \cite{LiJ12prb,LawKT09prl} with its transmission and reflection
amplitudes tuned by the Josephson phase $\phi$.

More generally, when $\mu\ne0$, the global transformation that block-diagonalizes
either partial Hamiltonian $H_{p}$ is not readily available, such
that we need to begin with generic scattering matrices at the QAHI-TSC
interfaces. To make progress, we use symmetry analysis and reduction
(see Sec. II B in Ref. \cite{Note-on-SM}). The strategy here is the
same as in Ref.~\cite{LiJ12prb}: By exploiting the particle-hole
symmetry and the electronic $U(1)$ gauge degree of freedom, we can
reduce a generic scattering matrix to its canonical, yet still general,
form which contains only symmetry-compliant and physically relevant
parameters. This leads to formally the same scattering matrices as
in Eqs.~(\ref{eq:Sp}-\ref{eq:SM}) except that: first, the Majorana
basis here is no longer attached to any (globally) block-diagonalized
Hamiltonians; second, $\varphi_{U}$ and $\varphi_{L}$ in general
become independent, such that the expressions for $t$ and $r$ in
Eq.~\eqref{eq:SM} become $t=({\cos\varphi_{U}+\cos\varphi_{L}})/({1+\cos\varphi_{U}\cos\varphi_{L}})$
and $r=-{\sin\varphi_{U}\sin\varphi_{L}}/({1+\cos\varphi_{U}\cos\varphi_{L}})$.
Indeed, by considering two limiting cases, with $\mu=0$ or $\phi=0$,
respectively, it is straightforward to deduce $\varphi_{U/L}=\pm\phi/2+k_{F}l_{C}$,
where $k_{F}$ is the Fermi wave vector for the QAHI edge mode in region
C. For the Hamiltonian in Eq.~\eqref{eq:H_BdG}, $k_{F}=\mu/v$.
Physically, this means that the propagation of a QAHI edge mode at
a finite momentum effectively introduces precession of Majorana polarization
to the composing $\chi$MMs \cite{Note-on-MP}. Finally, we obtain
the Majorana transmission and reflection amplitudes,
\begin{align}
t=\frac{2\cos(k_{F}l_{C})\cos({\phi}/{2})}{\cos^{2}(k_{F}l_{C})+\cos^{2}({\phi}/{2})},\;r=\frac{\cos^{2}(k_{F}l_{C})-\cos^{2}({\phi}/{2})}{\cos^{2}(k_{F}l_{C})+\cos^{2}({\phi}/{2})},
\end{align}
which are to be substituted into Eq.~\eqref{eq:SM}.

By using the scattering theory developed for $\chi$MM interferometry
in Ref.~\cite{LiJ12prb}, we further write down the average current
and the zero-frequency zero-temperature noise (shot noise) power in
the two normal contacts,
\begin{align}
 & I_{n}=\frac{e^{2}}{h}(1-r)V_{n},\qquad(n=1,2),\label{eq:In}\\
 & I_{s}=-(I_{1}+I_{2})=-\frac{e^{2}}{h}(1-r)(V_{1}+V_{2}),\label{eq:Is}\\
 & P_{a}=\frac{e^{3}}{h}\frac{t^{2}}{2}\max(|V_{1}|,|V_{2}|),\label{eq:Pa}\\
 & P_{c}=\frac{e^{3}}{h}\frac{t^{2}}{2}\text{sgn}(V_{1}V_{2})\min(|V_{1}|,|V_{2}|),\label{eq:Pc}
\end{align}
where $V_{n}$ is the bias voltage applied on contact $n=1,2$; $I_{n}\equiv\langle\hat{I}_{n}\rangle$
is the time-averaged current through normal contact $n$; $I_{s}$
is the time-averaged total current through the superconducting contacts
to the ground; $P_{a}\equiv P_{11}=P_{22}$ and $P_{c}\equiv P_{12}=P_{21}$
are the auto-correlator and the cross-correlator, respectively; $P_{nn'}\equiv\int_{-\infty}^{\infty}dt\,\frac{1}{2}\langle\{\hat{I}_{n}(t)-{I}_{n},\,\hat{I}_{n'}(0)-{I}_{n'}\}\rangle$,
with $\{$,$\}$ standing for the anti-commutator, is the zero-frequency
current correlation function (noise power) between normal contact
$n$ and $n'$ \cite{Buttiker92prb,Blanter00pr}. Several remarks
are in order. First, the average current $I_{n}$ in contact $n$
appears only depending on $r$ and $V_{n}$ on the same contact $n$.
This is a common feature of $\chi$MM interferometers resulting from
the fact that electric current can always be interpreted as interference
between two $\chi$MMs \cite{Note-on-CMM} \textendash{} only Majoranas
sourced from the same contact can maintain quantum phase coherence
in single-particle scattering processes, and therefore contribute
to a nonvanishing average current. Second, the current correlation
functions, in contrast to the average current, generally depend on
$t$, and the bias voltages on both normal contacts. In particular,
the current cross-correlator $P_{c}$ is contributed solely by the
exchange of two Majoranas sourced from different contacts in the form
of identical particles \cite{LiJ12prb}. Third, if measurements are
made by setting $I_{s}=0$, then the bias voltages must satisfy $V_{1}=-V_{2}$.
In this case, we denote $V_{0}=V_{1}-V_{2}=2V_{1}$, and $I_{0}=I_{1}=-I_{2}$.
From Eq.~\eqref{eq:In} we immediately obtain the two-terminal conductance,
$G\equiv I_{0}/V_{0}$, to be
\begin{align}
G_{\textrm{long}}=\frac{e^{2}}{h}\frac{\cos^{2}({\phi}/{2})}{\cos^{2}(k_{F}l_{C})+\cos^{2}({\phi}/{2})},\label{eq:Glong}
\end{align}
Clearly, $G_{\textrm{long}}$ oscillates with both $\phi$ and $k_{F}l_{C}$
as a consequence of the interference effect.

\textit{The short-junction limit.}\textemdash When the separation
$l_{C}$ between regions B and D becomes comparable to or less than
$\xi_{\mathrm{QAH}}$, the otherwise well-separated $\chi$MMs along
the B-C and the C-D interfaces strongly hybridize to become Andreev
bound states. The spectrum of these Andreev bound states is generally
gapped unless the Josephson phase $\phi\mod2\pi=\pi$ \cite{FuL08prl,FuL09prb}.
In this case, it is necessary to take into account the finite width
$W$ of any realistic sample, and hence the finite tunneling rate
of $\chi$MMs between the upper and the lower edges through the interface,
especially when the gap of the Andreev bound state spectrum approaches
0. In the following, we demonstrate the generic behavior of the MJI
in the short-junction limit by assuming $\mu=0$ and $l_{C}=0$ for
simplicity.

At the interface between the two TSCs (regions B and D with $l_{C}=0$),
the Andreev bound state dispersion can be solved at low energy to
be (see Sec. III A in Ref. \cite{Note-on-SM})
\begin{align}
E_{\text{Andreev}}\simeq\pm\sqrt{(vk_{y})^{2}+(\varepsilon\delta\phi)^{2}},
\end{align}
where $\varepsilon=(\Delta_{0}^{2}-m^{2})/(2\Delta_{0})$ and $\delta\phi\equiv(\phi-\pi)\mod2\pi$.
This indicates the gap along the interface varies as $|\varepsilon\delta\phi|$
when $\delta\phi$ is small, and the penetration depth along $\hat{y}$
of the evanescent states at $E=0$ is $\xi_{\phi}=v/|\varepsilon\delta\phi|$.
When $\xi_{\phi}\gtrsim W$, the tunneling of $\chi$MMs between the
upper and the lower edge of the sample along the interface becomes
significant \cite{Wieder14}. Such a tunneling problem can be explicitly solved in
the form of an effective model for the $\chi$MMs (see Sec. III B
in Ref. \cite{Note-on-SM}), which leads to the scattering relation
(cf. Eq.~\eqref{eq:SM})
\begin{align}
\begin{pmatrix}b_{E}^{(+)}\\
b_{A}^{(+)}
\end{pmatrix}=\begin{pmatrix}\tanh(W/\xi_{\phi}) & -\text{sech}(W/\xi_{\phi})\\
\text{sech}(W/\xi_{\phi}) & \tanh(W/\xi_{\phi})
\end{pmatrix}\begin{pmatrix}a_{A}^{(+)}\\
a_{E}^{(+)}
\end{pmatrix}.
\end{align}
Subsequently we obtain the average current and the current correlators
by substituting $t=\tanh(W/\xi_{\phi})$ and $r=\text{sech}(W/\xi_{\phi})$
into Eqs.~(\ref{eq:In}-\ref{eq:Pc}). In particular, by setting
$I_{s}=0$, the two-terminal conductance becomes
\begin{align}
G_{\textrm{short}}\simeq\frac{e^{2}}{h}\frac{1-\text{sech}(\varepsilon W\delta\phi/v)}{2},\label{eq:Gshort}
\end{align}
which vanishes when $\delta\phi=0$ and saturates to $e^{2}/2h$ when
$\delta\phi\gg v/\varepsilon W$. Incidentally, we note that in the
short-junction limit of the MJI, the topological property of region
C becomes irrelevant as long as the interface $\chi$MMs couple strongly
to form Andreev bound states.

\begin{figure}
\includegraphics[width=1\linewidth]{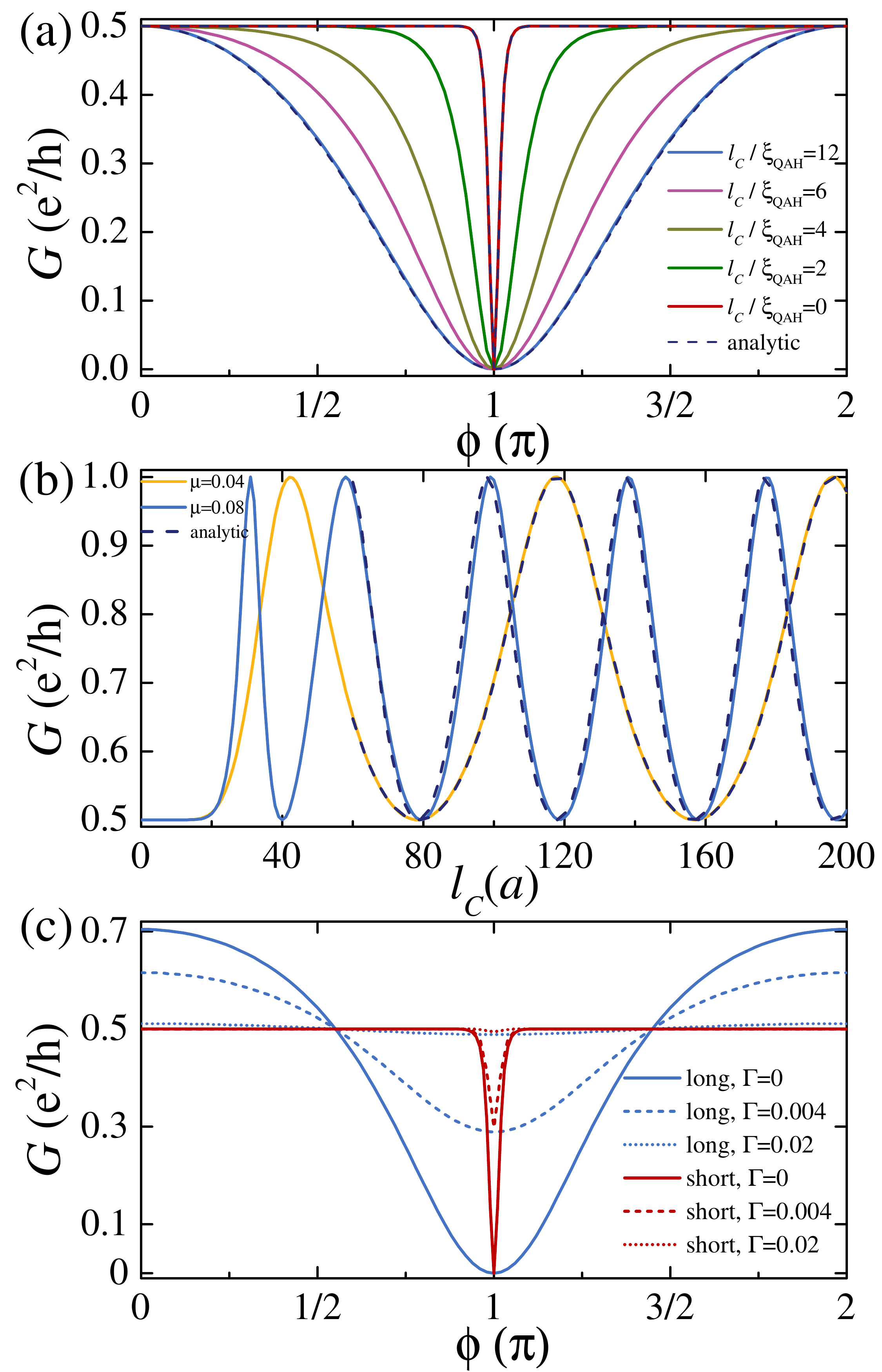} \caption{(a) Two-terminal conductance $G$ as a function of $\phi$ for different
junction length $l_{C}$ with the chemical potential $\mu=0$. The
solid lines are numerical results, whereas the dashed lines are analytical
results for the long-junction and the short-junction limits, respectively.
(b) $G$ as a function of $l_{C}$ with different $\mu$ at $\phi=0$.
The solid lines are numerical results, whereas the dashed lines are
analytical results for the long-junction limit plotted from $l_{C}=60a$.
(c) $G$ as a function of $\phi$ for different inverse quasiparticle
life time $\Gamma$ with $\mu=0.04$, in the long-junction (blue line,
$l_{C}=100a$) and the short-junction (red line) limits, respectively.
All plots here share the following parameters: $W=200a$, $x_{2}-x_{1}=x_{4}-x_{3}=100a$,
$b=v=1$, $\Delta_{0}=3m=0.3$. With these parameters we estimate
$\xi_{\mathrm{QAH}}/a\approx v/(2m)=5$. \label{fig:Conductance}}
\end{figure}

\textit{Numerical simulations.}\textemdash{} Up to now we have analyzed
two limits of the MJI to demonstrate its generic transport behavior,
as highlighted in Fig.~\ref{fig:Interfereometer}(b). In order to
extend our results to general settings beyond the analytically tractable
ones, we perform numerical simulations based on the discretized version
(with the lattice constant $a=1$) of the Hamiltonian in Eq.~\eqref{eq:H_BdG},
by using the Landauer-B$\ddot{\mathrm{u}}$ttiker formalism \cite{Buttiker92prb,Landauer70pm,Buttiker86prl,Fisher81prb,Dattabook}
adapted to superconducting systems \cite{Anantram96prb}. First of
all, we verify our preceding analytical results in Eqs.~\eqref{eq:Glong}
and \eqref{eq:Gshort}). The dependence of the numerically calculated
two-terminal conductance $G$ (see Sec. IV A in Ref. \cite{Note-on-SM})
on the Josephson phase $\phi$ and the junction length $l_{C}$ is
shown in Figs.~\ref{fig:Conductance}(a) and (b), respectively. We
find very good agreement between our numerical and analytical results
in both the long-junction and the short-junction limits. Next, we examine
the effect of inelastic scattering in the form of a finite quasiparticle
life time \cite{Ji18prl,HuangY18prb}, signified by $1/\Gamma$ (see
Sec. IV C in Ref. \cite{Note-on-SM}). Evidently shown in Fig.~\ref{fig:Conductance}(c),
the interference pattern weakens with increasing $\Gamma$, and disappears
when $G$ becomes a constant $e^{2}/2h$ at large enough $\Gamma$
\cite{Ji18prl,HuangY18prb}.

\textit{Discussion.}\textemdash One obvious advantage of the MJI is
that its interference pattern is a direct manifestation of phase coherent
$\chi$MM transport, and hence can be used as a smoking-gun signature
for the presence of $\chi$MMs in the setup. This will solve the current
controversy over the origin of the half-quantized conductance plateau
in the experiment reported in Ref.~\cite{He17sci} \textendash{}
the trivial mechanisms such as those proposed in Refs.~\cite{Ji18prl,HuangY18prb}
generally rely on substantial electron inelastic scattering especially
around the half-quantized plateau region, which necessarily destroys
the interference pattern. As such, the MJI can also be used as a tool
to measure the decoherence rate of the $\chi$MMs caused by various
environmental noises. Here, we stress that the physics of the interference effect
in an MJI is intrinsically \textit{different} from that of the well-established dc Josephson effect: The former concerns the \textit{nonequilibrium} current carried by the $\chi$MMs and measured at the \textit{normal metallic} contacts; the latter concerns the \textit{equilibrium} supercurrent through the \textit{superconducting} contacts that is not necessarily associated with any $\chi$MMs \cite{olund_current-phase_2012, snelder_andreev_2013, chen_emergent_2018}. More importantly, the MJI may further offer an effective platform for the braiding of chiral
Majorana fermions \cite{LianB17arXiv}, or even the manipulation of
Majorana qubits that can be defined upon the Majorana zero modes induced
in the vortices in the TSC regions \cite{Beenakker18arXiv}. An in-depth
investigation in this direction will be the subject of our future work.
\begin{acknowledgments}
C.A.L. thanks Bo Fu for helpful discussions. C.A.L. and S.Q.S. were partially
supported by the Research Grants Council, University Grants Committee,
Hong Kong under Grant No. 17301717 and C6026-16W. J.L. acknowledges
support from National Natural Science Foundation of China under Project
11774317. C.A.L. also acknowledges WIAS for hospitality where part of this work was carried out.
\end{acknowledgments}


\begin{thebibliography}{10}
\bibitem{Note-on-MP}The origin of Majorana polarization is transparent
and is associated with the electronic $U(1)$ degree of freedom in
defining Majorana modes: $\gamma_{+}(\phi)=\frac{1}{\sqrt{2}}(e^{i\phi/2}c+e^{-i\phi/2}c^{\dagger})$
and $\gamma_{-}(\phi)=\frac{i}{\sqrt{2}}(e^{i\phi/2}c-e^{-i\phi/2}c^{\dagger})$.
\bibitem{Note-on-CMM}As an intuitive way to see the interpretation
of electric current as interference between two $\chi$MMs, we simply
notice that the electron number operator can be expressed in terms
of the multiplication of two Majorana operators: if $\gamma_{+}=\frac{1}{\sqrt{2}}(c+c^{\dagger})$
and $\gamma_{-}=\frac{i}{\sqrt{2}}(c-c^{\dagger})$, then $\hat{N}=c^{\dagger}c=\frac{1}{2}(1+i\gamma_{+}\gamma_{-})$.
\end{thebibliography}
\end{document}